\documentclass[
reprint,
superscriptaddress,
amsmath,
amssymb,
aps,
pra,
]{revtex4-2}

\usepackage{graphicx}
\usepackage{dcolumn}
\usepackage{bm}
\usepackage{threeparttable}
\usepackage{tablefootnote}
\usepackage{color}
\usepackage{siunitx}
\usepackage{comment}
\newcommand{\rev}[1]{{\color{black} #1}}
\usepackage{lineno}

\usepackage{float}

\begin{document}
\preprint{APS/123-QED}

\title{Direct observation of photon-induced vortices in superconducting films}

\author{Takeshi Jodoi}
\address{National Institute of Advanced Industrial Science and Technology, 1-1-1, Umezono, Tsukuba, Ibaraki 305-8563, Japan}
\address{School of Engineering, The University of Tokyo, 2-11-16, Yayoi, Bunkyo, Tokyo 113-8656, Japan}

\author{Fuminori Hirayama}
\address{National Institute of Advanced Industrial Science and Technology, 1-1-1, Umezono, Tsukuba, Ibaraki 305-8563, Japan}

\author{Tetsuya Tsuruta}
\address{National Institute of Advanced Industrial Science and Technology, 1-1-1, Umezono, Tsukuba, Ibaraki 305-8563, Japan}

\author{Takahiro Kikuchi}
\address{National Institute of Advanced Industrial Science and Technology, 1-1-1, Umezono, Tsukuba, Ibaraki 305-8563, Japan}

\author{Daiji Fukuda}
\address{National Institute of Advanced Industrial Science and Technology, 1-1-1, Umezono, Tsukuba, Ibaraki 305-8563, Japan}




\date{\today}
\preprint{Submitted to Physical Review Applied}

\begin{abstract}
\rev{
Nucleation of vortex--antivortex pairs (VAPs) is believed to play a central role in the photon detection mechanism of superconducting detectors; 
however, their direct dynamic observation has remained challenging. 
Here, we report the direct observation of photon-induced VAP dynamics in a current-carrying superconductor as quantized voltage signals following photon absorption. 
The observed signals are interpreted as discrete phase-slip events, where each vortex traversal induces a $2\pi$ phase change of the superconducting order parameter, 
resulting in a quantized voltage pulse whose time integral is given by the magnetic flux quantum $\Phi_0$. 
We analyze the resulting quantized signals as a function of bias current, base temperature, and input photon-number states, 
and find that the number of VAPs generated per absorbed photon becomes effectively stabilized under specific conditions. 
Under these conditions, we demonstrate photon-number-resolving capability by directly counting phase-slip-induced voltage quanta. 
Our results reveal a detection mechanism governed by phase dynamics rather than conventional resistive transitions. 
We further show that photon-number resolution emerges when the fluctuation of photon-induced vortex–antivortex pair generation becomes statistically suppressed. 
These findings establish a new route toward photon-number-resolving detection based on phase-slip counting and open opportunities for high-speed superconducting 
detectors for quantum optics and photonic quantum technologies.
}
\end{abstract}

\maketitle


\section{Introduction}
\rev{The detection of single photons using superconducting devices is a key technology for quantum information, sensing, and optical communication. Conventional superconducting photon detectors, 
such as transition-edge sensors (TESs)~\cite{irwin2005transition,ullom2015review} and superconducting nanowire single-photon detectors 
(SNSPDs)~\cite{gol2001picosecond,natarajan2012superconducting,reddy2020superconducting},
rely on photon-induced suppression of superconductivity and the resulting resistive response.}

\rev{In these detectors, photon absorption creates a nonequilibrium region in which the superconducting order parameter is locally suppressed. This process is commonly described by hotspot-based models, 
including early theoretical descriptions~\cite{semenov2001quantum,semenov2005spectral} and subsequent refinements and analyses~\cite{zotova2012photon,vodolazov2014current, renema2013universal,hofherr2010intrinsic}.
Under a bias current, such nonequilibrium regions facilitate the generation of multiple vortex--antivortex pairs (VAPs), whose motion leads to Joule heating and can trigger the formation of a normal conducting domain.}

\rev{In particular, vortex-based descriptions have been proposed, where photon absorption induces vortex--antivortex pair (VAP) nucleation through the temporal suppression and recovery of the superconducting
 order parameter~\cite{engel2015detection,zotova2012photon,semenov2008vortex}. 
 The number of such events is expected to depend on the ratio between the strip width and the superconducting coherence length~\cite{zotova2012photon}. 
 In conventional SNSPDs with relatively narrow strips, multiple VAPs are generated, and their motion results in the formation of a resistive region. The detection signal is therefore governed 
 by the resulting resistive transition, which depends on the bias current, thermal relaxation, and circuit conditions.}

\begin{figure}[t]
    \centering
    \includegraphics[width=1.0\columnwidth]{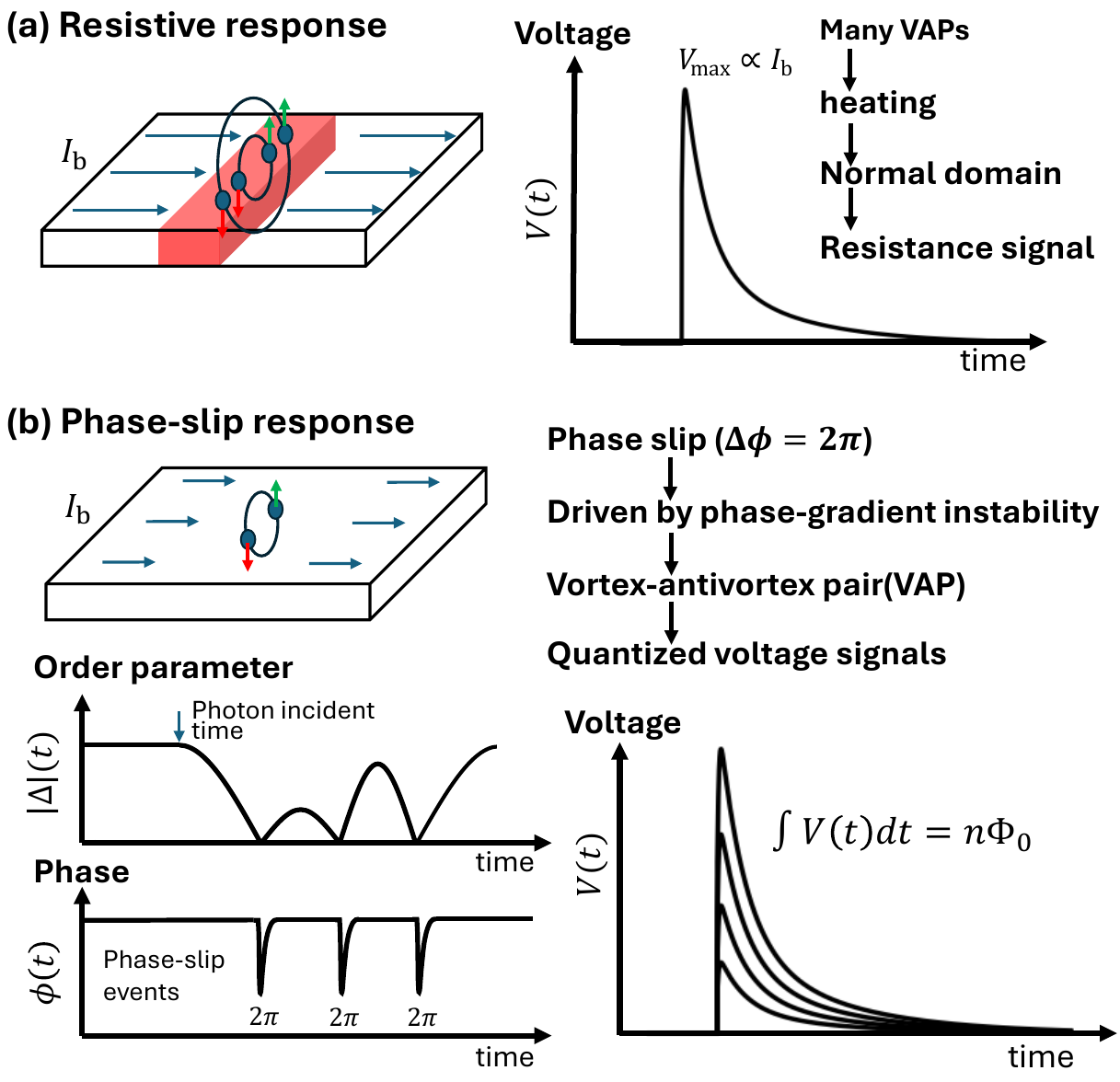}
    \caption{\rev{Schematic illustration of photon detection mechanisms: 
    (a) conventional resistive detection, in which photon absorption generates many vortex--antivortex pairs, resulting in heating, normal-domain formation, and a resistive response;
    and (b) VAP-mediated phase-slip detection (VBD in this work), in which only a few VAPs are generated and directly detected as quantized voltage pulses. The lower panels illustrate the transient suppression and recovery of the superconducting order parameter $\Delta(t)$ and the corresponding phase evolution $\phi(t)$ associated with discrete phase-slip events.}}
    \label{fig:detection-mechanism}
\end{figure}

 \rev{In addition, photon-number-resolving capabilities have also been actively explored in superconducting nanowire detectors~\cite{mattioli2015photon}.}
\rev{In contrast to these conventional approaches, the present work focuses on the nonequilibrium phase dynamics associated with photon absorption in a current-carrying superconductor. 
Rather than relying on the formation of a macroscopic resistive state, 
we investigate the regime in which only a small number of vortex--antivortex pairs are generated and their dynamics can be directly resolved.
The conceptual difference between these two detection mechanisms is illustrated schematically in Fig.~\ref{fig:detection-mechanism}}.

\rev{Photon absorption locally suppresses the superconducting order parameter, forming a transient nonequilibrium region (hotspot-like region). 
This region should not be confused with the normal core of a vortex; rather, it represents a temporary reduction of the superconducting condensate. 
Under bias current, the nonuniform current distribution leads to an instability in the phase gradient of the superconducting order parameter. 
This instability is released through the nucleation of vortex--antivortex pairs, which corresponds to discrete phase-slip events.}

\rev{When a vortex traverses the superconducting strip, the superconducting phase undergoes a $2\pi$ change. 
According to the Josephson relation,}
\[
\rev{V = \frac{\hbar}{2e} \frac{d\phi}{dt},}
\]
\rev{each phase-slip event produces a voltage pulse whose time integral satisfies}
\[
\rev{\int V(t)\,dt = \Phi_0,}
\]
\rev{where $\Phi_0 = h/2e$ is the flux quantum. Thus, the observed voltage pulses directly reflect the underlying phase evolution associated with vortex motion.}

\rev{In this work, we demonstrate the direct observation of photon-induced vortex--antivortex dynamics through quantized voltage signals. 
By resolving the number of phase-slip events occurring after photon absorption, 
we show that photon-number information does not automatically emerge from quantized phase-slip signals. 
Instead, it appears only when the fluctuation of the VAP generation process becomes sufficiently suppressed. 
Under these conditions, the VAP statistics approach the photon statistics of the incident light, enabling photon-number resolution through direct phase-slip counting.
}

\rev{The key distinction of this approach is that the detection signal is governed by discrete phase-slip events rather than by a continuous resistive transition. 
While the peak amplitude of the voltage signal depends on the readout circuit and bias current, the time-integrated voltage is determined solely by the phase change and remains quantized in units of $\Phi_0$. 
This allows direct access to the number of phase-slip events induced by photon absorption.}
\rev{In the following, we refer to this device as a vortex-based detector (VBD).}

\rev{The remainder of this paper is organized as follows. We first describe the device structure and measurement setup, 
and then present experimental observations of quantized voltage pulses. 
Finally, we analyze the dependence of the phase-slip dynamics on bias current, temperature, and photon number, and discuss the implications for photon-number-resolving detection.}

\section{Experiment}
\subsection{Experimental Methods}
The VBD used in the experiment consisted of a bilayer superconducting film composed of titanium and gold with thicknesses of 20\,nm and 10\,nm, respectively. 
This bilayer film was patterned into a square strip measuring \SI{8}{\micro m} $\times$ \SI{8}{\micro m}. The substrate of this device was a Si wafer. To enhance the detection efficiency, a Ti (5 nm)/Au (50 nm)/Ti (5 nm) was sputtered to form an Au mirror, and an $\rm{SiO_2}$ layer was subsequently deposited on top of the Au mirror to electrically insulate the active area. The titanium layers in this trilayer served as an adhesion layer. The Ti/Au bilayer was then deposited on the $\rm{SiO_2}$ layer, and an additional five-layer $\rm{SiO_2}/\rm{Si_3N_4}$ stacks were sputtered to form an anti-reflection layer. Niobium superconducting leads were fabricated on opposite sides of the square to read the VBD response signal. 
The superconducting properties of the VBD were measured in a dilution refrigerator and are summarized in Table~\ref{tab:performance_etftes}. 
\rev{The device performance is comparable to previously reported low-$T_\mathrm{c}$ optical TES devices~\cite{hattori2022optical}.}

Based on the measured superconducting parameters, we estimated $\xi_{\mathrm{GL}}(0) = 62.5~\mathrm{nm}$ for $T_c = 128~\mathrm{mK}$ and $D = 1.67~\mathrm{cm}^2/\mathrm{s}$~\cite{hook2013solid,irwin2005transition}. 
\rev{The coherence length was not directly measured from magnetic-field-dependent measurements, but estimated from the diffusion constant using standard relations.} 
\rev{Thus, the ratio $w / \xi_{\mathrm{GL}}(0) \approx 128$ for our VBD. 
In contrast to conventional narrow SNSPDs, where rapid formation of extended resistive regions can dominate the response depending on the bias current, 
this large ratio places the device in a regime in which only a limited number of vortex--antivortex pairs are generated, 
enabling direct observation of discrete phase-slip events.}

\begin{figure}[htpb]
    \centering
    \includegraphics[width = 1.0\columnwidth]{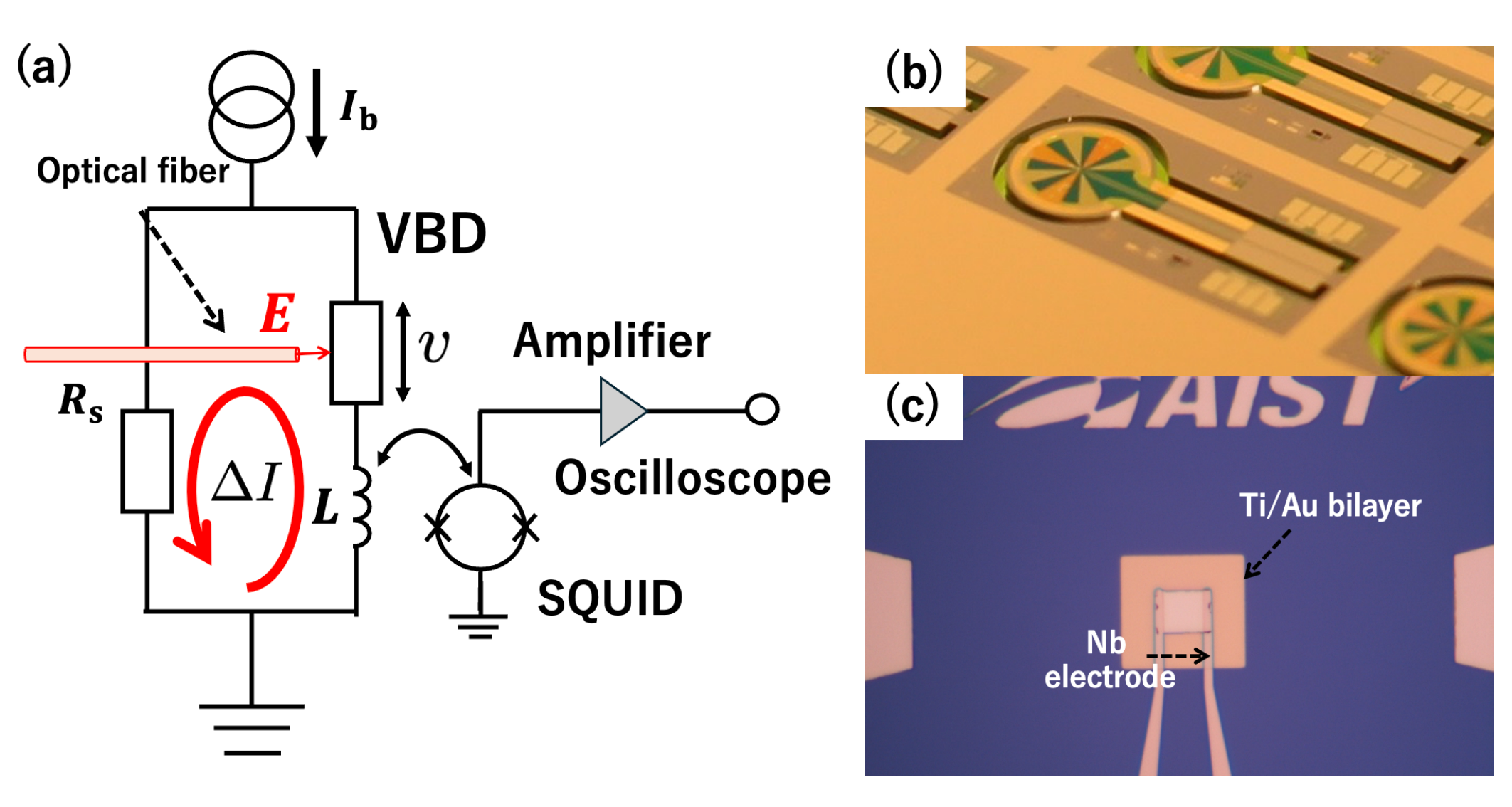}
    \caption{Readout circuit for the VBD. The rectangle with an arrow represents the VBD as the photon detection device. $R_\mathrm{s}$: shunt resistance; $I_\mathrm{b}$: bias current; $L$: kinetic inductance; $\Delta{I}$: change in current flowing in this readout circuit. (b) The VBD is fabricated on a key-shaped Si substrate to improve alignment between the optical fiber and the VBD. (c) The active area of the VBD. This active area consists of the Ti (20 nm)/Au (10 nm) bilayer, and the electrodes are made of niobium.}
    \label{fig:readoutcircuit}
\end{figure}

Figure~\ref{fig:readoutcircuit}(a) shows the readout circuit for the VBD, where the VBD and an inductor with inductance $L$ are shunted by a resistor $R_\mathrm{s}$. 
A bias current $I_\mathrm{b}$ is supplied to the VBD from an external current source. The VBD is cooled to a base temperature $T_\mathrm{b} < T_\mathrm{c}$ so that under equilibrium conditions, almost all of $I_\mathrm{b}$ flows through the VBD. The VBD is optically coupled to a fiber, 
and a laser pulse with an average photon number $\mu_{\mathrm{in}}$ is delivered to it through the fiber. 
\rev{The wavelength of the incident photons was \SI{1526}{nm}.}
After photon absorption in the VBD, a voltage signal $\Delta v(t)$ results from either the formation of resistance or the induced electromotive force due to the VAP motion, which is read out by the current amplifier of a superconducting quantum interference device (SQUID). From the observed current change $\Delta I(t)$ at the SQUID output, we obtain
\[
    \Delta v(t) = R_\mathrm{s} \times \Delta I(t)
\]
with a bandwidth of $0.35 / \tau_\mathrm{el}$, where we define the electrical time constant $\tau_\mathrm{el} = L / R_\mathrm{s}$.

\rev{For reference, the device can also exhibit a resistive response similar to electrothermal-feedback-based TES detection. However, in the present work we focus on a regime where the device remains superconducting and the detection signal arises from discrete phase-slip events associated with vortex motion. Figure~\ref{fig:readoutcircuit}(b) illustrates the VBD structure fabricated on a key-shaped Si substrate, and Fig.~\ref{fig:readoutcircuit}(c) shows the Ti/Au active area of the VBD together with the niobium electrodes. Table~\ref{tab:performance_etftes} summarizes representative ETF-TES-mode performance metrics for reference.}
\rev{To determine the appropriate operating conditions for the VBD mode, in particular the bias current relative to the critical current, 
we first investigated the dependence of the critical current $I_\mathrm{c}$ on the base temperature $T_\mathrm{b}$ by applying a triangular-shaped current to the VBD. 
The critical current $I_\mathrm{c}$ was defined as the current at which the resistance of the VBD exceeded $3~\mathrm{m\Omega}$.}
The temperature dependence of $I_\mathrm{c}$ is shown in Fig.~\ref{fig:Icorsignal}(a).
At lower values of $T_\mathrm{b}$, $I_\mathrm{c}(T)$ is almost constant. 
\rev{We evaluated $I_\mathrm{c}(0)$ = 7.36~\si{\micro A} by extrapolating the data obtained in the range $30~\mathrm{mK} < T_\mathrm{b} < 50~\mathrm{mK}$.} 
$I_\mathrm{c}(T)$ gradually vanishes as $T_\mathrm{b}$ approaches $T_\mathrm{c}$.
\begin{table}[htbp]
\caption{Performance of the superconducting device in the ETF-TES mode.}
 \label{tab:performance_etftes}
\begin{ruledtabular}
\begin{tabular}{lcl}
Material & &Ti (20 nm)/Au (10 nm) \\
Size & & \SI{8}{\micro m} $\times$ \SI{8}{\micro m} \\
Critical temperature &$T_\mathrm{c}$ & 128~\si{mK} \\
Effective time constant &$\tau_\mathrm{eff}$ & \\
&    & \quad 4.82~\si{\micro s}  ($T_\mathrm{b}=6$ mK)\\
&    & \quad 29.1~\si{\micro s} ($T_\mathrm{b}=117$ mK) \\
Normal resistance &$R_\mathrm{n}$ & 2.87~\si{\ohm} \\
Detection efficiency &$\eta$ & 76~\% \\
Critical current & $I_\mathrm{c}(T \to 0)$ & 7.36~\si{\micro A} (extrapolated)
\end{tabular}
\end{ruledtabular}
\end{table}
\subsection{Photon Response of the VBD}
\begin{figure*}[htbp]
    \centering
    \includegraphics[width = 1.9\columnwidth]{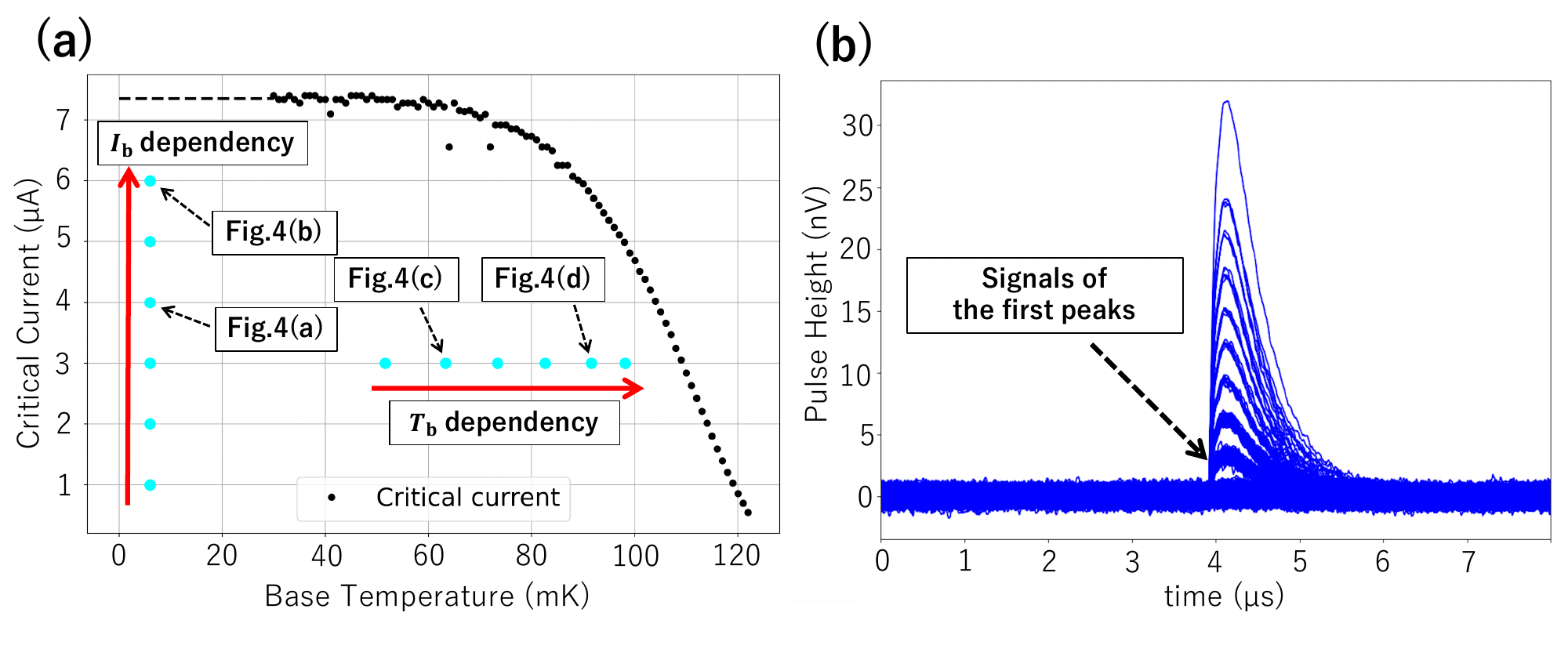}
    \caption{(a) Experimental data showing the dependence of the critical current on the base temperature. 
    These experimental data of the critical current are obtained from the current--voltage characteristics, and the dotted line is extrapolated to determine $I_\mathrm{c}(T = 0\,\rm{mK})$. 
    (b) Recorded signals at $I_{\rm{b}}=$\SI{5.0}{\micro A}, $T_{\rm{b}}=$ \SI{6}{mK}, $R_{\rm{s}} = 21.4$ $\mathrm{m\Omega}$, and $\mu_{\rm{in}} = 10$ photons per pulse. Under this condition, the time constant of the VBD is $\tau = 445$ ns.}
    \label{fig:Icorsignal}
\end{figure*}

To evaluate the photon response, we irradiated the VBD with photons while applying a bias current $I_\mathrm{b} < I_\mathrm{c}$, ensuring that the device remained in its superconducting equilibrium state. This condition closely resembles the standard operating mode of SSPDs. 
Figure$~\ref{fig:Icorsignal}$(b) shows the signals observed under $I_\mathrm{b} = 5.0$ \si{\micro A}, 
$T_\mathrm{b} = 6~\mathrm{mK}$, $R_\mathrm{s} = 21.4~\mathrm{m\Omega}$, and $\mu_{\mathrm{in}} = 10$ photons/pulse. 
\rev{Here, $\mu_{\mathrm{in}}$ denotes the average number of incident photons per pulse.}
\rev{The corresponding average number of detected photons is given by $\eta \mu_{\mathrm{in}}$, where $\eta$ is the detection efficiency.}
The pulse heights are quantized into discrete voltage steps in response to the incident laser pulses, a behavior fundamentally different from that of SSPDs. 

The measured time constant was $445~\mathrm{ns}$, approximately ten times faster than that in the ETF-TES mode (see Table~I). 
To clarify whether quantization originates from resistance or voltage changes, we measured the dependence of $\Delta v$ on $I_\mathrm{b}$ and $T_\mathrm{b}$. 
If $\Delta v$ is caused by a resistance change $\Delta R$, it will scale proportionally with $I_\mathrm{b}$ because $\Delta R$ remains nearly constant at a fixed $T_\mathrm{b}$. Likewise, for a constant $I_\mathrm{b}$, $\Delta v$ will vary with $T_\mathrm{b}$ because the heat capacity of the VBD changes with the temperature $T_\mathrm{b}$. 

\begin{figure*}[htpb]
    \centering
    \includegraphics[width = 1.9\columnwidth]{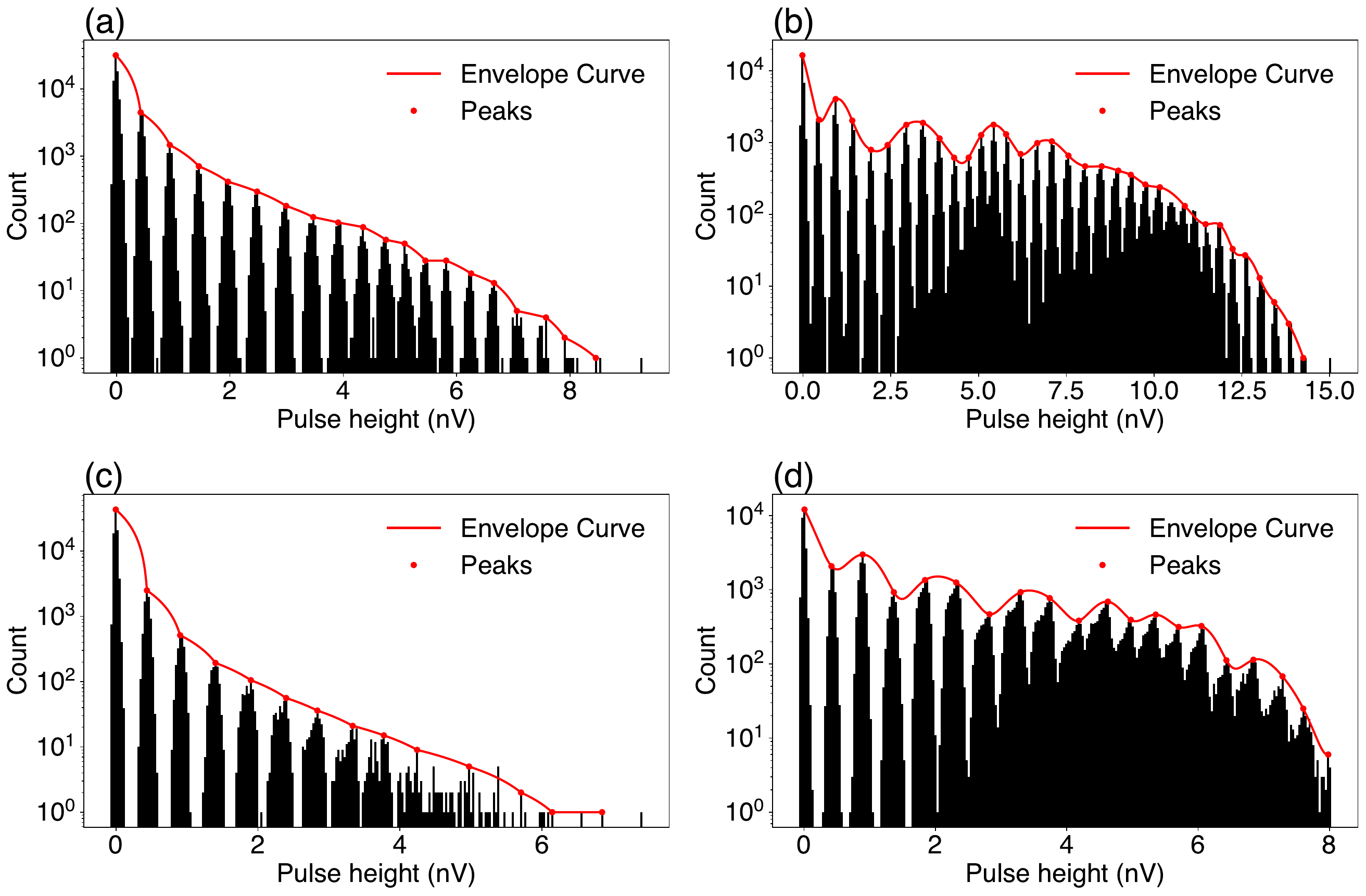}
    \caption{Pulse-height distributions under four different experimental conditions, where the peak positions are marked.  
    \rev{The envelope curves are included only as visual guides for peak evolution and do not represent a theoretical model.}
    (a) $I_{\rm b}=$\SI{4.0}{\micro A}, $T_{\rm b}=6~\mathrm{mK}$,  $\mu_{\rm in}=5$ photons per pulse;
    (b) $I_{\rm b}=$\SI{6.0}{\micro A}, $T_{\rm b}=6~\mathrm{mK}$,  $\mu_{\rm in}=5$ photons per pulse;
    (c) $I_{\rm b}=$\SI{3.0}{\micro A}, $T _{\rm b}=63~\mathrm{mK}$,$\mu_{\rm in}=5$ photons per pulse;
    (d) $I_{\rm b}=$\SI{3.0}{\micro A}, $T_{\rm b}=91~\mathrm{mK}$, $\mu_{\rm in}=5$ photons per pulse.
    \rev{In all cases, discrete peaks are observed, where each peak corresponds to the number of phase-slip events. The peak structure depends on both bias current $I_{\rm b}$ and base temperature $T_{\rm b}$, indicating that the number of generated VAPs varies systematically with the operating conditions.} }
    \label{fig:pulse_height}
\end{figure*}

Figure$~\ref{fig:pulse_height}$ shows the pulse-height distribution for $\mu_{\mathrm{in}} = 5$ photons/pulse, including markers for the peak positions and an overlaid envelope curve. 
Figure$~\ref{fig:pulse_height}$(a) and $\ref{fig:pulse_height}$(b), obtained at $T_\mathrm{b} = 6~\mathrm{mK}$, demonstrate that increasing $I_\mathrm{b}$ from \SI{4}{\micro A} to \SI{6}{\micro A} produces additional peaks, even at the same photon input. Similarly, for a constant $I_\mathrm{b}$, the \rev{peak index} increases as $T_\mathrm{b}$ rises from $63~\mathrm{mK}$ to $91~\mathrm{mK}$. Across all conditions, the voltage pulses remain quantized at intervals of approximately $0.45~\mathrm{nV}$, forming discrete peaks regardless of $I_\mathrm{b}$ or $T_\mathrm{b}$. 

\rev{In Fig.~\ref{fig:pulse_height}(d), pulses with amplitudes exceeding 8 nV are not observed. 
When the pulse height at single peak is 0.45 nV, the current diverted to the shunt resistor is approximately \SI{0.2}{\micro A}. 
Thus, the number of pulse peaks is determined by the bias current. Therefore, the pulse heights in Fig.~\ref{fig:pulse_height}(d) are truncated at around 8 nV because the $I_\mathrm{b}=$ \SI{3}{\micro A}. 
Furthermore, the minimum peak amplitude in Fig.~\ref{fig:Icorsignal}(b) is approximately an order of magnitude smaller than that in Fig.~\ref{fig:pulse_height}. 
This difference is attributed to the change in the electrical time constant resulting from the different value of $R_\mathrm{s}$. }
These results indicate that photon detection in a VBD is governed by a mechanism fundamentally distinct from that of resistance variation. 
Notably, Fig.~\ref{fig:pulse_height}(b) and \ref{fig:pulse_height}(d), where $I_\mathrm{b}$ or $T_\mathrm{b}$ is close to $I_\mathrm{c}$ or $T_\mathrm{c}$, respectively, 
exhibit periodic modulations of the peak counts. These modulations are evident from the envelope formed by connecting the peak maxima in the figure.
\rev{This behavior indicates enhanced fluctuations in the number of VAP events under these operating conditions, suggesting that the VAP generation process is not fully stabilized and motivating the statistical analysis presented in Section~III.}

\begin{figure*}[t]
    \centering
    \includegraphics[width = 1.9\columnwidth]{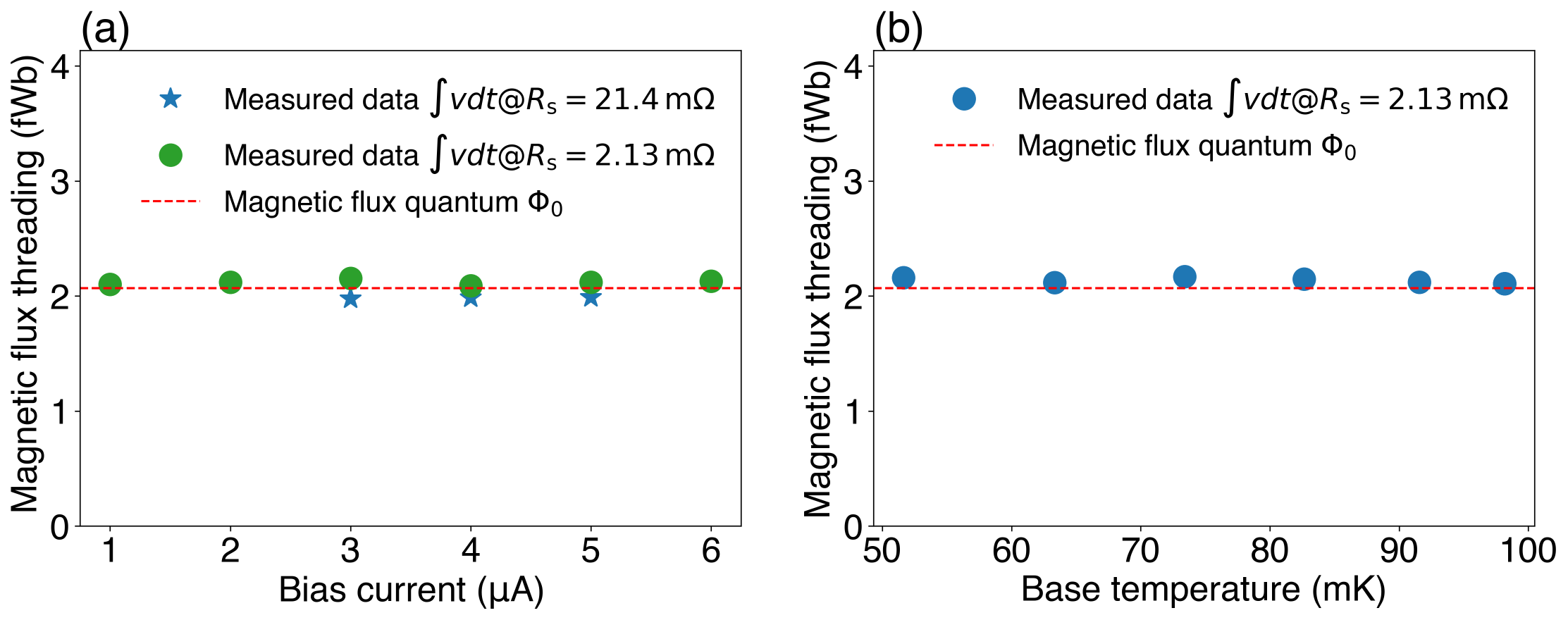}
    \caption{(a) Relationship between voltage--time integral at each bias current and magnetic flux quantum. We obtained these data from $I_\mathrm{b}=$\SI{1}{\micro A} to \SI{6}{\micro A} at $T_\mathrm{b}=$\SI{6}{mK}, following the $I_\mathrm{b}$ dependency in Fig.~\ref{fig:Icorsignal}(a). (b) Relationship between voltage--time integral at each base temperature and magnetic flux quantum.
     The data was obtained from $T_\mathrm{b}=$ \SI{52}{mK} to \SI{98}{mK} (\SI{52}{mK}, \SI{63}{mK}, \SI{73}{mK}, \SI{83}{mK}, \SI{92}{mK}, \SI{98}{mK}) at $I_\mathrm{b}=$\SI{3}{\micro A}, following the $T_\mathrm{b}$ dependency in Fig.~\ref{fig:Icorsignal}(a). At $T_\mathrm{b}$ = 98 mK, the voltage--time integral is calculated from the time constant of the first stage in the averaged pulse.}
    \label{fig:phi_dependence}
\end{figure*}

\section{Photon detection mechanism based on phase-slip dynamics}
\subsection{Quantization of Signal Waveforms in Units of Magnetic Flux Quantum}
As discussed in the previous section, \rev{the observed signal in the VBD mode does not primarily originate from the generation of resistance following photon absorption because the minimum pulse integral remains independent of both $I_\mathrm{b}$ and $T_\mathrm{b}$}. \rev{Instead, we interpret the observed voltage in terms of the phase evolution of the superconducting order parameter. 
According to the Josephson relation,}
\begin{equation}
\rev{V(t) = \frac{\hbar}{2e}\frac{d\phi}{dt},}
\end{equation}
\rev{where $\phi$ is the superconducting phase. When a vortex traverses the superconducting strip, the phase changes by $2\pi$, and therefore the time-integrated voltage satisfies}
\begin{equation}
\rev{\int V(t)\,dt = \frac{\hbar}{2e}\Delta\phi = n_\mathrm{VAP}\Phi_0,}\label{eqn:integral_voltage_signal}
\end{equation}
\rev{where $n_\mathrm{VAP}$ denotes the number of phase-slip events (or equivalently the number of traversing VAPs) and $\Phi_0=h/2e$ is the flux quantum.}

\rev{From this point of view, we obtained the voltage--time integral in Eq.~(\ref{eqn:integral_voltage_signal}) for the minimum-amplitude pulses under different experimental conditions involving $I_\mathrm{b}$ and $T_\mathrm{b}$. Figure~\ref{fig:phi_dependence}(a) and \ref{fig:phi_dependence}(b) show the dependence of this quantity on the bias current and base temperature, respectively. In Fig.~\ref{fig:pulse_height}, the voltage pulses remain quantized with a constant interval of approximately 0.45~nV. Therefore, we extracted the waveform corresponding to the first peak index and calculated its voltage--time integral to discuss its dependence on $I_\mathrm{b}$ and $T_\mathrm{b}$. At $T_\mathrm{b}=98$~mK, the first-peak statistics were insufficient, so we instead used the waveform corresponding to peak index 3 and divided the resulting integral by the peak index to compare with the single-event value.}

\rev{For all experimental conditions, the minimum pulse integral remains almost constant and close to the magnetic flux quantum $\Phi_0 = 2.07 \times 10^{-15}~\mathrm{Wb}$. The value does not depend on $R_\mathrm{s}$. These results show that the minimum-amplitude voltage pulse corresponds to a single phase-slip event, i.e., $n_\mathrm{VAP}=1$.}

Here, we assume that the quantized pulse shape originates from the quantum number $n_\mathrm{VAP}$ and examine the dependence of $\Phi$ on $n_\mathrm{VAP}$. 
Figure~6(a) shows the results under two different experimental conditions with respect to $I_\mathrm{b}$ and $T_\mathrm{b}$. Specifically, we define case 1 as $I_\mathrm{b}=$\SI{6}{\micro A} and $T_\mathrm{b}=6~\mathrm{mK}$, 
and case 2 as $I_\mathrm{b}=$\SI{3}{\micro A} and $T_\mathrm{b}=98~\mathrm{mK}$. 
For case 1, the value of $\Phi$ is proportional to $n_\mathrm{VAP}$ up to approximately 17, which clearly indicates that the observed signals in Fig.~3(b) are quantized according to $n_\mathrm{VAP}$. However, for $n_\mathrm{VAP} > 17$, this proportionality is no longer valid. Under the conditions of case~2, the value of $\Phi$ significantly exceeds $\Phi_0$ even for small values of $n_\mathrm{VAP}$. 
To investigate this, we focus on the pulse shape of the observed voltage signals. Figure~6(b) shows an example of the averaged pulse shape used for the decomposition analysis.
Apparently, the signal shape consists of two exponential components with distinct time constants, which is completely different from the shape shown in Fig.~3(b). 
\begin{figure*}[t]
    \centering
    \includegraphics[width = 2.0\columnwidth]{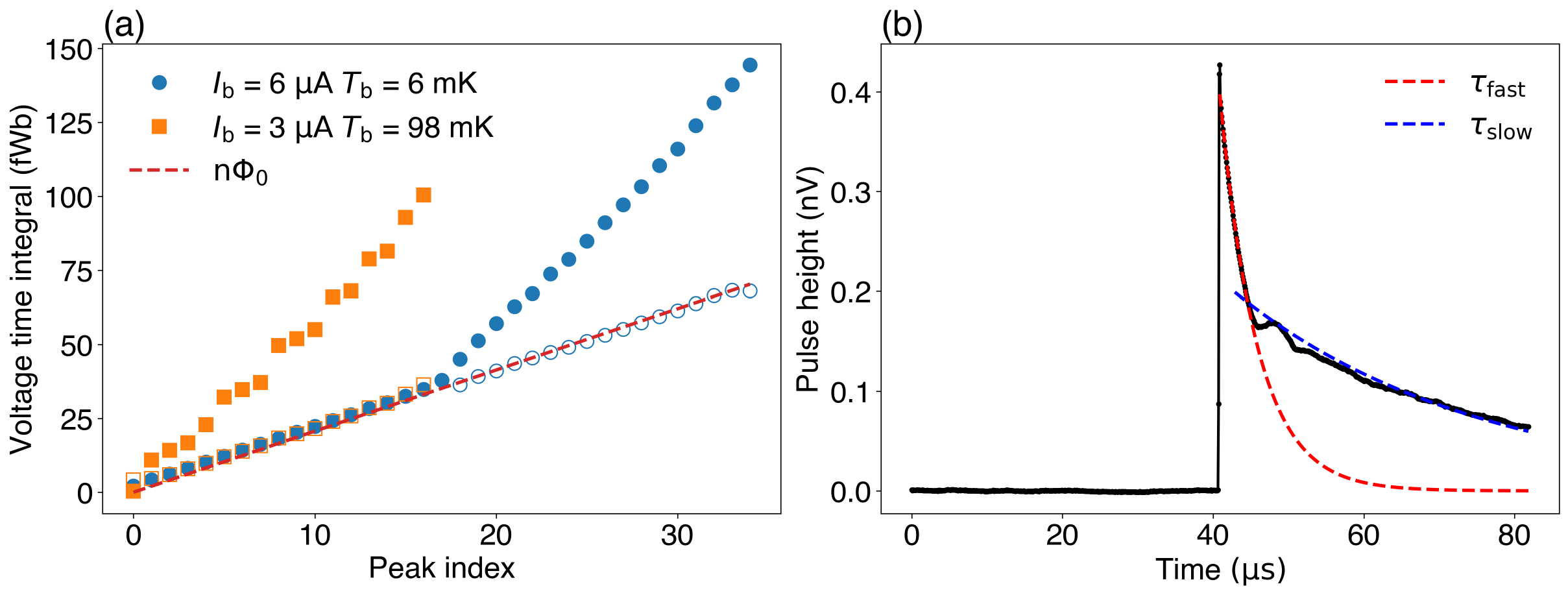}
    \caption{(a) Dependence of the voltage–time integral of individual pulse peaks on the peak index. Closed circles and rectangles represent $\int{v}dt$ of the obtained average pulse, whereas the results obtained by integrating the $\tau_\mathrm{fast}$ component of the separated signals are replotted as open circles or rectangles. The open symbols exhibit good consistency with the straight line having a slope of $\Phi_0$. (b) Averaged pulse obtained from the VBD at $T_{\rm{b}}$ = 98 mK. The time constant of this waveform appears to increase in the later part owing to Joule heating.}
    \label{fig:phi-v.s-peakindex}
\end{figure*}

To separate the two components, we fit the signal with
\begin{equation}
f(t) = a \exp(-t/\tau_\mathrm{fast}) + b \exp(-t/\tau_\mathrm{slow}), \label{eqn:pulse_shape_fitting}
\end{equation}
which gives $a = 0.397~\mathrm{nV}$, $b = 0.207~\mathrm{nV}$, $\tau_\mathrm{fast} =$\SI{5.06}{\micro s}, and $\tau_\mathrm{slow} =$\SI{32.4}{\micro s}.
We performed this separation for the other signals for each $n_\mathrm{VAP}$, 
and the results obtained by integrating the $\tau_\mathrm{fast}$ component of the separated signals are replotted in Fig.~6(a) as open circles or rectangles.

In this case, all the values are proportional to $\Phi_0$ with $n_\mathrm{VAP}$, which shows that the first component of Eq.~(\ref{eqn:pulse_shape_fitting}) can 
be attributed to the traversal of the magnetic flux quanta, consistent with the expected phase-slip dynamics.

In addition, $\tau_\mathrm{slow}$ in the slow component of Eq.~(\ref{eqn:pulse_shape_fitting}) is close to the ETF-TES time constant $\tau_\mathrm{ETF} = 29.1~\si{\micro s}$,
measured at $T_\mathrm{b} = 117~\mathrm{mK}$, as shown in Table~\ref{tab:performance_etftes}. 
\rev{In the ETF-TES operation, the effective time constant increases as the base temperature approaches the critical temperature.}
Therefore, the slow component can be attributed to the generation of a resistive change in the transition region after the traversal of the magnetic flux quanta.

\begin{figure*}[htpb]
    \centering
    \includegraphics[width = 2.0\columnwidth]{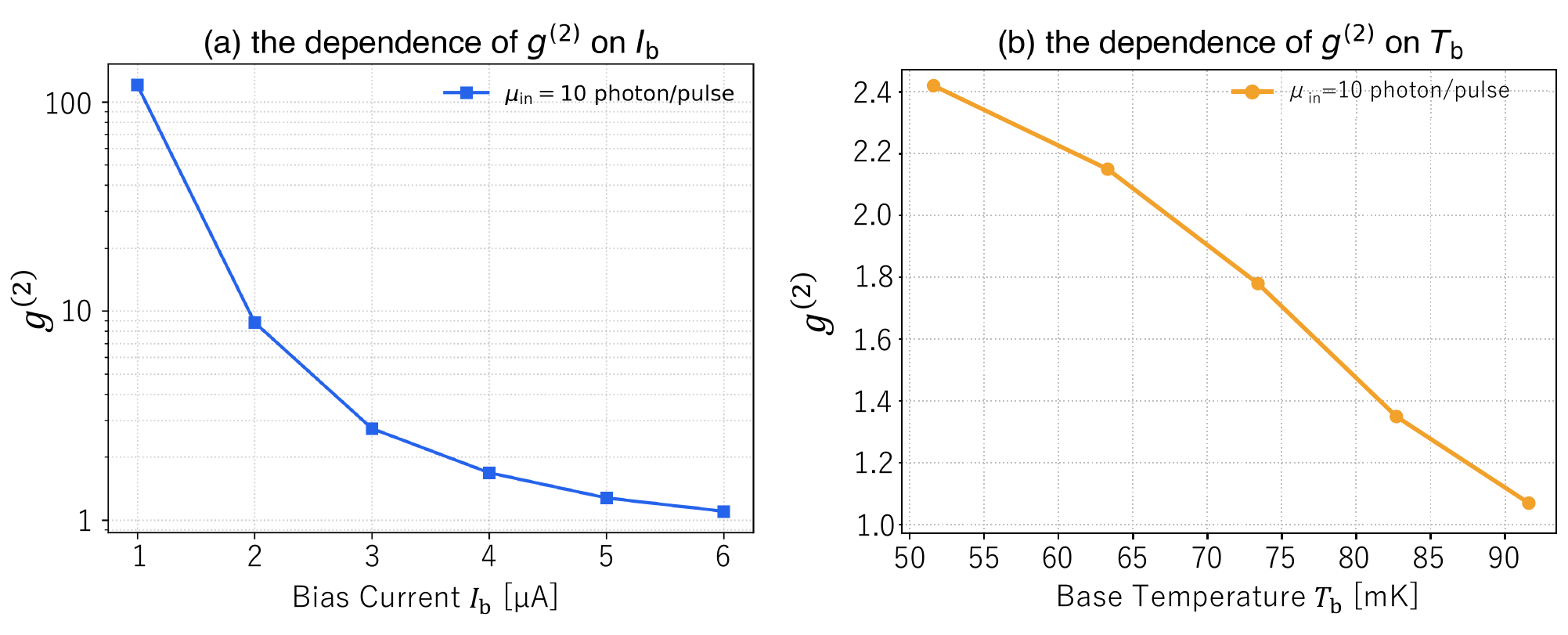}
    \caption{\rev{(a) Dependence of the second-order correlation function $g^{(2)}_{\mathrm{VAP}}$, calculated from the number of observed vortex–antivortex (VAP) events per pulse, on the bias current $I_{\rm b}$ 
    at a base temperature of $T_{\rm b} = 97$~mK. 
    (b) Dependence of $g^{(2)}_{\mathrm{VAP}}$ on the base temperature $T_{\rm b}$ at a fixed bias current of $I_{\rm b} = \SI{3}{\micro A}$. 
    In both cases, $g^{(2)}_{\mathrm{VAP}}$ approaches unity with increasing $I_{\rm b}$ or $T_{\rm b}$, corresponding to the Poisson limit. 
    These results indicate that fluctuations in the number of VAP events per pulse are suppressed, identifying the operating conditions under which 
    the VAP generation process becomes effectively stabilized. }}
    \label{fig:g2-dependency}
\end{figure*}

\subsection{Signal Response Speed}
In the VBD operation mode, the signal response time is in the order of $\tau \sim w / v$, where $w$ and $v$ denote the strip width of the superconducting film and traversing velocity of the VAPs, respectively. The velocity $v$ for the titanium/gold superconducting film is not known; however, previous research reports that the velocity for an NbC superconductor is approximately $10^{4}~\mathrm{m/s}$ \cite{dobrovolskiy2020ultra}.
Assuming this estimation holds for the TiAu film, the signal response time can be approximated as $\tau \sim 800~\mathrm{ps}$ for $w =$ \SI{8}{\micro m}. Consequently, the VBD mode is expected to exhibit a substantially faster response when compared with that of the ETF mode. Nevertheless, the observed time constant in Fig.~3(b) is restricted to $\tau = 445~\mathrm{ns}$. This limitation arises from the bandwidth of the bias circuit depicted in Fig.~2, where the signal response is governed by the electrical time constant $\tau_{\mathrm{elc}} = L / R_\mathrm{s}$. We estimated $L = 9.8~\mathrm{nH}$ by measuring the noise spectrum of the VBD, which yields $\tau_{\mathrm{elc}} = 458~\mathrm{ns}$ for $R_\mathrm{s}=21.4~\mathrm{m\Omega}$.
This value is in good agreement with the time constant of the observed signals. 
Therefore, a wide-bandwidth readout is crucial for achieving an intrinsically faster signal response in the VBD.

\subsection{PNR Capability}

As discussed in the previous section, the response signal is quantized \rev{according to the number of phase-slip events. This quantization enables photon-number resolution
by directly counting vortex–antivortex (VAP) events.}
\rev{To quantitatively evaluate this behavior, we introduce the average number
of traversing VAPs per absorbed photon, denoted as $n_{\mathrm{VAP}}$,
which refers to the average number of traversing VAPs per absorbed photon
as defined in Eq.~(\ref{eqn:n_VAP}). This quantity provides a measure of the
conversion between absorbed photons and the resulting phase-slip events.}

The average number of traversing VAPs is obtained from the observed distribution of phase-slip events using the following equation:
\begin{equation}
\rev{n_{\mathrm{VAP}}=\frac{1}{\eta \, {\mu}_{\mathrm{in}}}\sum_{n=0} n \, P(n), }\label{eqn:n_VAP}
\end{equation}
where $P(n)$ represents the probability of observing $n$ magnetic flux quanta, and $\eta$ is the detection efficiency. 

\rev{To identify the operating conditions under which fluctuations of $n_{\mathrm{VAP}}$ are suppressed, we evaluate the second-order correlation function $g^{(2)}_{\mathrm{VAP}}$, defined from the number of VAP events observed per pulse, as a function of bias current and base temperature,
as shown in Fig.~\ref{fig:g2-dependency}. 

Figure~\ref{fig:g2-dependency}(a) shows the dependence of $g^{(2)}_{\mathrm{VAP}}$ 
on the bias current at $T_b=$ \SI{97}{mK},
while Fig.~\ref{fig:g2-dependency}(b) shows its dependence on the base temperature
at $I_b=$ \SI{3}{\micro A}.
We find that $g^{(2)}_{\mathrm{VAP}}$ approaches unity
under specific combinations of $I_b$ and $T_b$,
indicating that fluctuations in the number of VAP events per pulse are suppressed.
\rev{Using these operating conditions, we examine the photon-number-resolving (PNR) behavior under
$T_{\mathrm{b}} = 98~\mathrm{mK}$ and $I_{\mathrm{b}}=$ \SI{3}{\micro A}, with
${\mu}_{\mathrm{in}} = 1~\text{photon/pulse}$.}}

\begin{figure*}[htpb]
    \centering
    \includegraphics[width = 2.0\columnwidth]{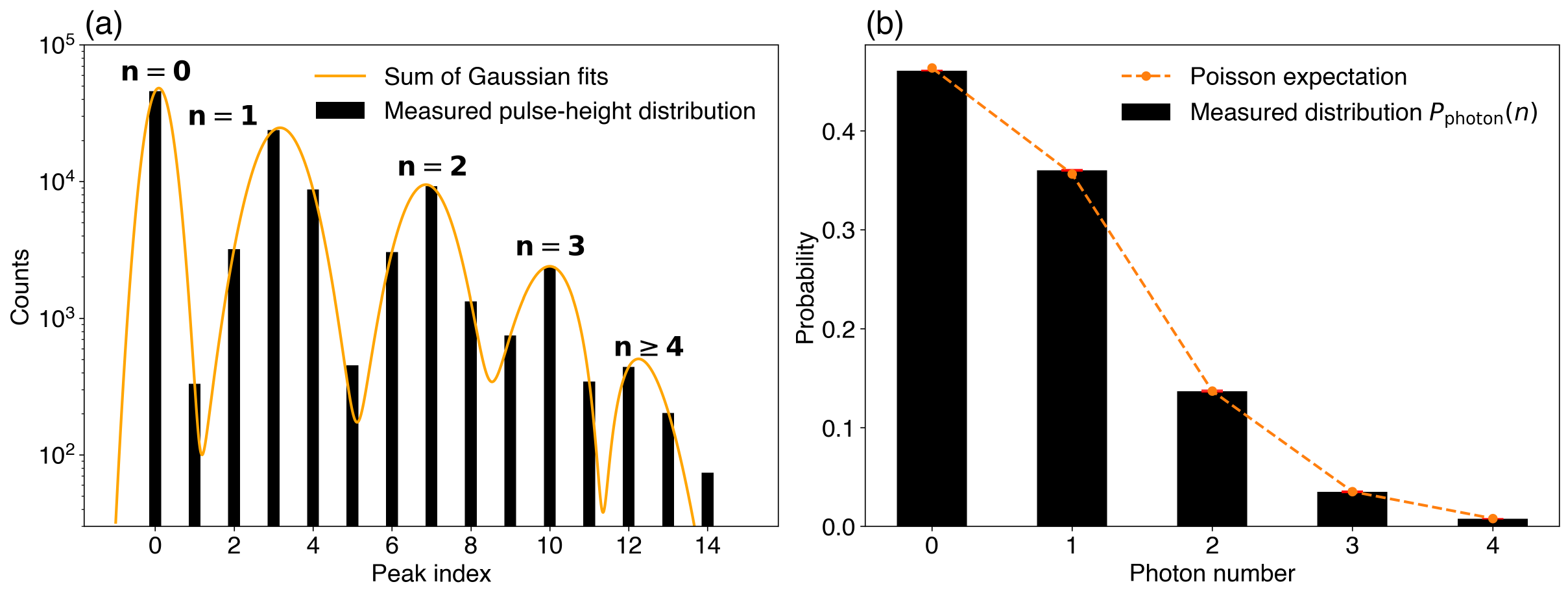}
    \caption{\rev{Photon-number-resolving behavior measured under operating conditions 
    identified in Fig.~7, where $g^{(2)}_{\mathrm{VAP}} \approx 1$ (at $T_{\rm b} = 97$~mK), indicating stabilized VAP generation.
    (a) Distribution of the number of observed VAP events per pulse.
    (b) Corresponding photon-number distribution inferred from the VAP counts. 
    Under these conditions, both distributions follow Poisson statistics. 
    The second-order correlation function evaluated from the photon-number 
    distribution, $g^{(2)}_{\mathrm{photon}} \approx 1$, is consistent with Poissonian photon statistics.
    These results demonstrate that the Poissonian nature of the VAP statistics 
    ($g^{(2)}_{\mathrm{VAP}} \approx 1$) is directly transferred to the photon-number statistics, enabling photon-number-resolving behavior.}}
    \label{fig:PoissonStatistics}
\end{figure*}

Under these conditions, we directly evaluate the statistical distributions of the detected events to verify the emergence of photon-number-resolving behavior.
The bars in Fig.~\ref{fig:PoissonStatistics}(a) show the distribution of the individual magnetic quantum peaks under these conditions. The total number of events is $N_{\mathrm{all}} = 10^5$. 
From this distribution, we obtained $n_{\mathrm{VAP}}$ using Eq.~(\ref{eqn:n_VAP}), which yields $ n_{\mathrm{VAP}}=3.22~\Phi_0/\text{photon}.$

In addition to this method, we applied a multi-Gaussian fitting to the distribution to determine $n_{\mathrm{VAP}}$. The fitting function is expressed as

\begin{equation}
f(x) = \frac{N_\mathrm{all}}{\sqrt{2\pi}} \sum_{n=0} \frac{P_\mathrm{photon}(n)}{\sigma_n} \exp\left[-\frac{(x - b_n)^2}{2\sigma_n^2}\right],
\end{equation}
where $N_{\mathrm{all}}$ is the total number of events, $P_{\mathrm{photon}}(n)$ is the probability of the $n$-photon state, $b_n$ is the center position of the $n$th peak, and $\sigma_n$ is its standard deviation. 
After fitting, we obtained 
$n_{\mathrm{VAP}} = \sum b_n/\sum n= 3.18,$
which agrees well with the value obtained by the previous method in Eq.~(\ref{eqn:n_VAP}). 
The bars in Fig.~\ref{fig:PoissonStatistics}(b) show the probability of the photon-number states $P_\mathrm{photon}(n)$ determined by the fitting. 
The dotted line represents the Poisson distribution under the same condition, with an average photon number of $\eta\mu_\mathrm{in} = 0.77$. 
We conducted a chi-square test on the datasets, and the results revealed no significant differences among the conditions ($\chi^2 = 3.41$, $p = 0.333$). 

\rev{Moreover, the second-order correlation function
$g^{(2)}_{\mathrm{photon}}$ for photons was evaluated to be
$g^{(2)}_{\mathrm{photon}} = 0.972$, consistent with Poissonian photon statistics.
This result demonstrates that the Poissonian nature of the VAP statistics
($g^{(2)}_{\mathrm{VAP}} \approx 1$) is directly transferred to the
photon-number statistics.}
\rev{It should be noted that these measurements are performed under specific
operating conditions identified through the $g^{(2)}_{\mathrm{VAP}}$ analysis,
in contrast to the general response shown in Fig.~\ref{fig:pulse_height}, 
where the VAP number fluctuates and does not directly reflect photon statistics.}

\rev{To compare the experimentally obtained value with a simple energy-scale estimate, we introduce the kinetic-inductive energy scale associated with a $2\pi$ phase-slip event,}
\begin{equation}
\rev{E_{\Phi_0}=\frac{\Phi_0^2}{2L_\mathrm{k}},}
\end{equation}
\rev{where $L_\mathrm{k}$ is the kinetic inductance of our device. Using the estimated kinetic inductance under $T_\mathrm{b}=98$~mK, $I_\mathrm{b}=3~\mu$A, and an input wavelength of 1526~nm, we obtain $E_\gamma/E_{\Phi_0}\approx2.88$, which is in reasonable agreement with the experimentally obtained value. We emphasize that this estimate is not a rigorous prediction of the number of VAPs; rather, it provides an order-of-magnitude consistency check, since the actual number of events is governed by nonequilibrium phase-slip dynamics under the given bias and thermal conditions.}

The present work establishes the existence of a direct statistical link between absorbed-photon energy and vortex–antivortex generation. 
Future studies of wavelength dependence and dynamic range will provide further insight into the energy-conversion process from photons to phase-slip events.

\section{Conclusion}
In summary, we experimentally demonstrated the direct observation of photon-induced VAP dynamics in superconducting films operated in a vortex-based detection mode. 
\rev{By analyzing the quantized voltage signals associated with phase-slip events,} 
we confirmed that the observed response \rev{associated with phase-slip events (i.e., magnetic flux quanta traversing the superconducting strip)} rather than the resistance changes in the transition region. 
The voltage-time integrals exhibit clear quantization in units of the magnetic flux quantum, establishing a direct link between the signal amplitude and the number of VAPs. Furthermore, we identified that the measured response time is limited by the readout circuit bandwidth, suggesting that sub-nanosecond intrinsic timescales are achievable through circuit optimization.

Importantly, we demonstrated the PNR capability based on the correlation between the number of absorbed photons and number of traversing VAPs. 
The reconstructed photon-number distribution agrees with the Poisson statistics, confirming the feasibility of multi-photon resolution in this detection scheme. These findings represent the first experimental evidence of VAP-mediated photon detection and provide a pathway toward fast, high-efficiency, multi-photon-resolving superconducting detectors. 
Such detectors could play a crucial role in quantum optics and quantum information processing, where precise photon-number discrimination and ultra-fast response are essential.
\rev{These results demonstrate that photon detection can be governed by phase-slip dynamics rather than conventional resistive transitions.}
\\

\textit{Acknowledgments-}
This work was supported in part by the BRIDGE program, Cross-ministerial SIP, JSPS KAKENHI Grant Number 24K01374, and JST Moonshot R\&D Program Grant Number JPMJMS2064-6 and JPMJMS256I-4.\\
D. Fukuda contributed equally to this work. F. Hirayama, T. Tsuruta, and T. Kikuchi led the discussion and interpretation of the results.

\bibliographystyle{apsrev4-2}
\bibliography{apssamp}
\end{document}